\documentstyle[prd,aps,preprint]{revtex}

\renewcommand{\vec}[1]{\mbox{\boldmath $#1$}}
\newcommand{\barq}{\langle {\bar q}q\rangle}
\newcommand{\sumint}{\sum\nolimits_m\hspace{-0.78cm}\int}
\begin{document}

\draft
\tighten

\title{Chiral phase transition at high temperature\\
 in the QCD-like gauge theory}

\author{O.  Kiriyama\thanks{e-mail address: kiriyama@nucl.phys.tohoku.ac.jp}, M.  Maruyama and F.  Takagi}
\address{Department of Physics,
Tohoku University, Sendai 980--8578, Japan}
\date{\today}
\maketitle

\begin{abstract}
The chiral phase transition at high temperature is investigated using the effective potential in the framework of the QCD-like gauge theory with a variational approach. We have a second order phase transition at $T_c=136$MeV.
We also investigate numerically the temperature dependence of $\barq$, $f_\pi$ and $a_2(T)$(coefficient of the quadratic term in the effective potential) and estimate the critical exponents of these quantities.  
\end{abstract}

\pacs{PACS numbers: 11.10.Wx,11.15.Tk,11.38.-t,12.38.Lg}
\section{INTRODUCTION}
The properties of Quantum Chromodynamics(QCD) at finite temperature are a subject of general interest. 
At zero temperature and zero chemical potential, the chiral symmetry is spontaneously broken. On the other hand, we expect that the chiral symmetry will be restored at sufficiently high temperature and it may be realized in high energy heavy-ion collisions at RHIC and LHC. The restoration of the chiral symmetry is important also in physics of neutron stars and early universe.

Effective theories of QCD such as $\sigma$ model\cite{OGURE} and Nambu-Jona-Lasinio\cite{ASA} model or lattice QCD\cite{LAT} have been used in the study of the chiral phase transition and the related critical phenomena in QCD.
An effective potential and/or Schwinger-Dyson equation for quark propagator have been widely used to study the spontaneous chiral symmetry breaking and its restoration at finite temperature and/or finite chemical potential. 

In this paper we study the chiral phase transition at finite temperature in the chiral limit. 
We use the Cornwall-Jackiw-Tomboulis(CJT) type effective potential\cite{CJT} for a composite operator in the framework of the renormalization group improved gauge theory(the so called QCD-like gauge theory)[5,6,7].
We take into account a correct asymptotic behavior of the quark self energy and use the variational method in calculating the effective potential.
The present approach does not break the chiral symmetry explicitly in contrast to some previous studies of the effective theories of QCD.
Using the effective potential we study the critical temperature and order of the chiral phase transition. If the transition is second order, we can calculate the critical exponents.
Rajagopal and Wilczek have studied the second order phase transition using three dimensional $O(4)$ Heisenberg model\cite{WIL}.
However it is still an open question whether QCD with two massless quarks and $O(4)$ Heisenberg model belong to the same universality class or not.
The critical exponents have been also studied using lattice QCD\cite{LAT}.

Furthermore, in previous studies of the effective theories which respect the QCD[9,10,11,12] it is not obvious whether the critical exponents agree with those of the mean field approximation a la Landau or not.
We numerically study the critical exponents of the order parameter for the chiral phase transition $\langle {\bar q}q\rangle_T$, the pion decay constant at finite temperature $f_{\pi}(T)$ and $a_2(T)$ which is the coefficient of the quadratic term in the effective potential.

This paper is organized as follows. 
In Sec.2 we briefly review the effective potential for a composite operator in the QCD-like gauge theory. 
In Sec.3 we calculate the effective potential at finite temperature. 
In Sec.4 we present numerical results and determine the temperature dependence of $\barq$ and $f_{\pi}$. We also estimate the critical exponents. 
Sec.5 contains the summary and discussion.

\section{EFFECTIVE POTENTIAL AT ZERO TEMPERATURE}
In this paper we fix the mass scale by the condition $\Lambda_{QCD}=1$.
The CJT effective potential for the quark propagator takes the following form in Minkowski space:
\begin{eqnarray}
V[G,D]&=&\int\frac{d^4p}{(2\pi)^4i}~\mbox{tr}\left(\ln (S^{-1}(p)G(p))-S^{-1}(p)G(p)+1\right)\nonumber\\
&&-\frac{i}{2}\int\int\frac{d^4p}{(2\pi)^4i}~\frac{d^4q}{(2\pi)^4i}~g^2C_2\mbox{tr}\left(\gamma_{\mu}G(p)\gamma_{\nu}G(q)\right)D^{\mu\nu}(p-q),\label{eqn:vg}
\end{eqnarray}
where $C_2=(N_c^2-1)/(2N_c)$ is the quadratic Casimir operator for color $SU(N_c)$ group, $S(p)$ is the propagator of a massless quark, and $D^{\mu\nu}(p)$ is the gluon propagator in the Landau gauge:
\begin{eqnarray}
D^{\mu\nu}(p)=\frac{i}{p^2}(g^{\mu\nu}-\frac{p^{\mu}p^{\nu}}{p^2}),\nonumber
\end{eqnarray}
which is diagonal in the color space and ``tr'' refers to Dirac, flavor and color matrices.
An alternative effective potential for composite operator which is a variation of Eq.(\ref{eqn:vg}) was developed in Ref.\cite{BAR1}. Here we use the one presented above. 

In the Landau gauge and the one-loop approximation, we can expand the quark two-point function as follows
\begin{eqnarray}
iG^{-1}(p)=p\hspace{-0.19cm}/-(\Sigma_S(p^2)+i\gamma_5\Sigma_P(p^2)),\label{eqn:sp}
\end{eqnarray}
where $\Sigma_S(p)$ and $\Sigma_P(p)$ are the scalar part and the pseudo-scalar part of the quark dynamical mass function, respectively. Substituting (\ref{eqn:sp}) into (\ref{eqn:vg}) and using the running coupling in the two-loop level potential we obtain the following effective potential $V=V_1+V_2$ which satisfies the renormalization group equation after Wick rotation,  
\begin{eqnarray}
V_1&=&-2N_fN_c\int\frac{d^4p}{(2\pi)^4}~\ln\frac{\Sigma_S^2(p^2)+\Sigma_P^2(p^2)+p^2}{p^2}\nonumber\\
&&+4N_fN_c\int\frac{d^4p}{(2\pi)^4}~\frac{\Sigma_S^2(p^2)+\Sigma_P^2(p^2)}{\Sigma_S^2(p^2)+\Sigma_P^2(p^2)+p^2},\label{eqn:vz1}
\end{eqnarray}
\begin{eqnarray}
V_2 &=& -6N_fN_cC_2\int\int\frac{d^4p}{(2\pi)^4}~\frac{d^4q}{(2\pi)^4}~\frac{{\bar g}^2(p,q)}{(p-q)^2}\nonumber\\
&\times&\frac{\Sigma_S(p^2)\Sigma_S(q^2)+\Sigma_P(p^2)\Sigma_P(q^2)}{(\Sigma_S^2(p^2)+\Sigma_P^2(p^2)+p^2)(\Sigma_S^2(q^2)+\Sigma_P^2(q^2)+q^2)},\label{eqn:vz2}
\end{eqnarray}
where $N_f$ is the number of the light flavors.
This expression is our starting point.
The order parameter can be determined by the condition
\begin{eqnarray}
\frac{\delta V}{\delta \Sigma_S(p^2)}=0~~,~~\frac{\delta V}{\delta \Sigma_P(p^2)}=0.\nonumber
\end{eqnarray}
These extremum conditions for $V$ with respect to $\Sigma_S(p^2)$ and $\Sigma_P(p^2)$ are equivalent to the Schwinger-Dyson equations for the quark propagator in the improved ladder approximation.
We take $\Sigma_P(p^2)=0$ without loss of generality because of the chiral symmetry.
We use Higashijima-Miransky approximation[5,6] for the running coupling with an infrared modification\cite{HIG91}
\begin{eqnarray}
{\bar g}^2(p,q)\rightarrow{\bar e}^2(\max(p^2,q^2))\label{eqn:HMA},
\end{eqnarray}
where 
\begin{equation}
{\bar e}^2(p^2)=\frac{2\pi^2a}{\ln(p^2+p_R^2)}\label{eqn:coup}.
\end{equation}
Here, $a=24/(11N_c-2n_f)$, $n_f$ is the number of flavors which controls the running coupling, and $p_R$ is a parameter to regularize the divergence at $p=1(\Lambda_{QCD})$. 
The modified running coupling has the same asymptotic behavior as the QCD running coupling, while it gradually approaches a constant as $p^2$ decreases. The value of the running coupling at $p=0$ is $2\pi^2a/t_R$, where $t_R=\ln (p_R^2)$ is the infrared regularization parameter.

As ${\bar e}^2(\max(p^2,q^2))$ has no angle dependence in four-dimensional Euclidean space, Eqs.(\ref{eqn:vz1}) and (\ref{eqn:vz2}) are reduced to
\begin{eqnarray}
V_1[\Sigma_S] &=& 4N_fN_c\int\frac{d^4p}{(2\pi)^4}~\frac{\Sigma_S^2(p^2)}{\Sigma_S^2(p^2)+p^2}\nonumber\\&&-2N_fN_c\int\frac{d^4p}{(2\pi)^4}~\ln\frac{\Sigma_S^2(p^2)+p^2}{p^2},\label{eqn:v1}\\
V_2[\Sigma_S] &=& -12N_fN_cC_2\int\frac{d^4p}{(2\pi)^4}~\frac{{\bar e}^2(p^2)}{p^2}~\frac{\Sigma_S(p^2)}{\Sigma_S^2(p^2)+p^2}\nonumber\\
&&\times\int_{p^2>q^2}\frac{d^4q}{(2\pi)^4}~\frac{\Sigma_S(q^2)}{\Sigma_S^2(q^2)+q^2}.\label{eqn:v2}
\end{eqnarray}

In this paper we consider the case where the chiral symmetry is exact, i.e. renormalized current quark masses are zero.
We use the variational method to investigate the vacuum of the QCD-like theory and use the following form for $\Sigma_S(p)$ as a trial mass function\cite{HIG91}:
\begin{equation}
\Sigma_S(p^2)=-\langle {\bar q}q\rangle~\frac{2\pi^2a}{3(p^2+p_R^2)}(\ln (p^2+p_R^2))^{\frac{12}{33-2n_f}-1},\label{eqn:s2}
\end{equation}
where $\langle {\bar q}q\rangle$ is the renormalization group invariant vacuum expectation value of ${\bar q}q$.
This form was chosen so that its asymptotic form
\begin{eqnarray}
\Sigma_S(p^2) \rightarrow -\langle {\bar q}q\rangle~\frac{2\pi^2a}{3p^2}(\ln p^2)^{\frac{12}{33-2n_f}-1}.\nonumber
\end{eqnarray}
coincides with the one obtained by using the operator product expansion and the renormalization group equation\cite{POL76}. This form of dynamical mass function does not break the chiral symmetry explicitly, i.e. the axial vector current is conserved\footnote{We note, however, it was found\cite{MUN} that the chiral Ward-Takahashi identity is violated unless gluon momentum squared is used as the argument for the running coupling. The solution to recover the chiral Ward-Takahashi identity in the framework of the improved ladder approximation was given in Ref.\cite{KUGO92}}.

%
%
%  FINITE TEMPERATURE
%
%
\section{EFFECTIVE POTENTIAL AT FINITE TEMPERATURE}

We apply the imaginary time formalism\cite{JIK} to calculate the effective potential at finite temperature.
Since fermion fields obey an anti-periodic boundary condition, we have
\begin{eqnarray}
p_4 \rightarrow \omega_n=(2n+1)\pi T, \label{eqn:im1}
\end{eqnarray}
and the four-momentum integration is replaced with the summation over the Matsubara-frequency $\omega_n$ and the three-momentum integration
\begin{eqnarray}
\int\frac{d^4p}{(2\pi)^4}\rightarrow T\sum_{n=-\infty}^{\infty}\int\frac{d{\vec p}}{(2\pi)^3}\label{eqn:im2}
\end{eqnarray}
in Eqs.(\ref{eqn:v1}) and (\ref{eqn:v2}).
Now we have
\begin{eqnarray}
V_1&=&4N_fN_cT\sum_{n=-\infty}^{\infty}\int\frac{d{\vec p}}{(2\pi)^3}~\frac{\Sigma_S^2(\omega_n,{\vec p})}{\Sigma_S^2(\omega_n,{\vec p})+Q_n^2}\nonumber\\
&&-2N_fN_cT\sum_{n=-\infty}^{\infty}\int\frac{d{\vec p}}{(2\pi)^3}~\ln\left(1+\frac{\Sigma_S^2(\omega_n,{\vec p})}{Q_n^2}\right),\label{eqn:v11}
\end{eqnarray}
\begin{eqnarray}
V_2&=&-12N_fN_cC_2T\sum_{n=-\infty}^{\infty}\int\frac{d{\vec p}}{(2\pi)^3}~\frac{{\bar e}^2(\omega_n,{\vec p})}{Q_n^2}~\frac{\Sigma_S(\omega_n,{\vec p})}{\Sigma_S^2(\omega_n,{\vec p})+Q_n^2}\nonumber\\
&&\times T\sumint~~\frac{d{\vec q}}{(2\pi)^3}~\frac{\Sigma_S(\omega_m,{\vec q})}{\Sigma_S^2(\omega_m,{\vec q})+Q_m^2},\label{eqn:v12}
\end{eqnarray}
where $Q_n^2=\omega_n^2+{\vec p}^2$, $Q_m^2=\omega_m^2+{\vec q}^2$, and
\begin{eqnarray}
\sumint~~\frac{d{\vec q}}{(2\pi)^3}\nonumber
\end{eqnarray}
denotes the summation over $m$ and integration over ${\vec q}$ with the condition $Q_n^2 > Q_m^2$ and we have replaced $\bar{e}^2(p^2)$ and $\Sigma_S(p^2)$ with $\bar{e}^2(\omega_n,\vec{p})$ and $\Sigma_S(\omega_n,\vec{p})$, respectively. Their functional forms are given below. 

Let us now consider the choice for the running coupling and the quark dynamical mass function at finite temperature. We expect that the running coupling decreases as the temperature increases. Usually a coupling of the form like
\begin{eqnarray}
g^2(T)=\frac{2\pi^2a}{\ln(T^2/M^2)}\nonumber
\end{eqnarray}
is used.
However we use a running form 
\begin{eqnarray}
{\bar e}^2(\omega_n,\vec p)=\frac{2\pi^2a}{\ln(\omega_n^2+{\vec p}^2+p_R^2)},\label{eqn:fe}
\end{eqnarray}
as a natural extension of the $T=0$ case(Eq.(\ref{eqn:coup})).

As concerns the dynamical mass function, we use the following function
\begin{eqnarray}
\Sigma_S(\omega_n,{\vec p})=-\langle {\bar q}q\rangle\frac{2\pi^2a}{3(\omega_n^2+{\vec p}^{~2}+p_R^2)}\left(\ln(\omega_n^2+{\vec p}^{~2}+p_R^2)\right)^{\frac{12}{33-2n_f}-1},\label{eqn:fq}
\end{eqnarray}
by replacing $p^2$ with $\omega_n^2+\vec{p}^2$ also in Eq.(\ref{eqn:s2}).
Here, $\barq$ is the variational parameter of the effective potential and its value is determined from the minimum of the potential. 
Note that the chiral symmetry is not explicitly broken by the choice Eq.(\ref{eqn:fq}).

Using the effective potential in Eqs.(\ref{eqn:v11}) and (\ref{eqn:v12}) with the ansatz Eq.(\ref{eqn:fe}), Eq.(\ref{eqn:fq}) and changing the variables
\begin{eqnarray}
t&=&\ln(\omega_n^2+{\vec p}^2+p_R^2),\nonumber\\
t_n&=&\ln(\omega_n^2+p_R^2),\nonumber
\end{eqnarray}
we obtain
\begin{eqnarray}
V(\sigma,T)=V_1(\sigma,T)+V_2(\sigma,T),\nonumber
\end{eqnarray}
\begin{eqnarray}
V_1(\sigma,T)&=&\frac{4N_fN_cT\sigma^2}{(2\pi)^2}\sum_{n=-\infty}^{\infty}\int_{t_n}^{\infty}dt~\sqrt{e^t-e^{t_n}}~\frac{e^{-2t}}{t^{2-a}(1-e^{t_R-t})+\sigma^2e^{-3t}}\nonumber\\&&-\frac{2N_fN_cT}{(2\pi)^2}\sum_{n=-\infty}^{\infty}\int_{t_n}^{\infty}dt~e^t\sqrt{e^t-e^{t_n}}~\ln\left(1+\frac{\sigma^2 e^{-3t}}{t^{2-a}(1-e^{t_R-t})}\right),\label{eqn:fv1}
\end{eqnarray}
\begin{eqnarray}
V_2(\sigma,T)&=&-\frac{8N_fN_caT^2\sigma^2}{(2\pi)^2}\sum_{n=-\infty}^{\infty}\int_{t_n}^{\infty}dt~\frac{\sqrt{e^t-e^{t_n}}}{1-e^{t_R-t}}~\frac{t^{-a/2}e^{-2t}}{t^{2-a}(1-e^{t_R-t})+\sigma^2e^{-3t}}\nonumber\\
&&\times\sum_m {}^{'}\int_{t_m}^t du~\sqrt{e^{u}-e^{t_m}}~\frac{u^{1-a/2}e^{-u}}{u^{2-a}(1-e^{t_R-u})+\sigma^2e^{-3u}},
\end{eqnarray}
where $\sigma=2\pi^2a\barq/3$ is the rescaled order parameter.
The sum over $m$, $\sum_m {}^{'}$, is taken to an extent that $t>u$.

In order to determine the critical temperature $T_c$, we expand the effective potential as a power series of the order parameter with finite coefficients $a_{2n}(T)$
\begin{eqnarray}
V(\sigma,T)=a_2(T)\sigma^2+O(\sigma^4).
\end{eqnarray}
The coefficient $a_2(T)$ is calculated as follows 
\begin{eqnarray}
a_2(T)&=&\frac{dV}{d\sigma^2}\bigg{|}_{\sigma^2=0}\nonumber\\
&=&\frac{2N_fN_cT}{(2\pi)^2}\left[\sum_{n=-\infty}^{\infty}\int_{t_n}^{\infty}dt~\sqrt{e^t-e^{t_n}}~\frac{e^{-2t}}{t^{2-a}(1-e^{t_R-t})}\right.\nonumber\\
&&\left.-4Ta\sum_{n=-\infty}^{\infty}\int_{t_n}^{\infty}dt~\sqrt{e^t-e^{t_n}}~\frac{t^{a/2-2}e^{-2t}}{(1-e^{t_R-t})^2}\sum_m {}^{'}\int_{t_m}^t du~\sqrt{e^{u}-e^{t_m}}~\frac{u^{a/2-1}e^{-u}}{1-e^{t_R-u}}\right].
\end{eqnarray}
The critical temperature $T_c$ is determined by the condition, $a_2(T_c)=0$.

%
%
%
%   NUMERICAL RESULTS
%
%
%
\section{NUMERICAL RESULTS}
\subsection{The Behavior of the Effective Potential at Finite Temperature}
In our calculation we put $N_f=2$, $N_c=3$ and $n_f=3$. 
We use the three flavor($n_f=3$) running coupling since three flavors(u,d,s) contribute to the running coupling in the energy region of the chiral phase transition. 
Since it has been shown that the quantities such as $\barq$ and $f_\pi$ are not sensitive to the infrared regularization parameter $t_R$ at zero temperature\cite{HIG91}, we expect that these quantities are insensitive to $t_R$ also at finite temperature as long as $t_R$ is not too large.
In Fig.1 we plot the $t_R$ dependence of the value of $\barq$, the location of minimum, the pion decay constant $f_{\pi}$ which is approximately given by the Pagels-Stoker formula\cite{PS}:
\begin{eqnarray}
f_{\pi}^2=4N_c\int\frac{d^4p}{(2\pi)^4}~\frac{\Sigma_S(p^2)}{(\Sigma_S^2(p^2)+p^2)^2}\left(\Sigma_S(p^2)-\frac{p^2}{2}\frac{d\Sigma_S(p^2)}{dp^2}\right)\nonumber
\end{eqnarray}
and the constituent quark mass $m$ defined by the relation, $m=\Sigma_S(p^2=4m^2)$.
The fact that these values are insensitive to the infrared cutoff at zero temperature has been also pointed out by Kugo {\it et al}\cite{KUGO92} who used the Schwinger-Dyson equation in the improved ladder approximation.
Therefore we fix the infrared regularization parameter $t_R$ to $0.1$.
In Fig.2 we show that at zero temperature the approximate form of the dynamical mass function $\Sigma_S(p)$ at the minimum of $V$ well reproduces the exact solution of the Schwinger-Dyson equation. Therefore we adopt Eq.(\ref{eqn:fq}) as a trial function and use $\sigma$ as a variational parameter at finite temperature.
We have confirmed that the quantities such as $\barq_T$ and $f_\pi(T)$ are insensitive to the infrared regularization parameter $t_R$ at finite temperature.

The effective potential at finite temperature is shown in Fig.3. 
At zero temperature the chiral symmetry is spontaneously broken. As temperature grows the minimum of the potential continuously approaches zero. Hence we have second order phase transition at $T=0.185(\Lambda_{QCD})$. To determine the value of $\Lambda_{QCD}$ we use the Pagels-Stoker formula and fix the pion decay constant $f_{\pi}$ to $93$ MeV at zero temperature and have $\Lambda_{QCD}=737$MeV.
Then the restoration of the chiral symmetry occurs at $T_c=136$MeV.

The temperature dependence of the order parameter $\langle {\bar q}q\rangle_T$ is shown in Fig.4.
As concerns the temperature dependence of $\langle {\bar q}q\rangle_T$ 
we refer, for example, to that of the free pion gas approximation:
\begin{eqnarray}
\frac{\barq_T}{\barq_0}=1-\frac{T^2}{8f_{\pi}^2}.\label{eqn:cpt}
\end{eqnarray}
This behavior is also shown in Fig.4 and there is a slight difference between our result and Eq.(\ref{eqn:cpt}) in the low temperature region.

Once the quark dynamical mass function is determined from the minimum of the effective potential, we can calculate the pion decay constant $f_{\pi}$ at finite temperature using the Pagels-Stoker formula extended to finite temperature:
\begin{eqnarray}
f_{\pi}(T)^2=4N_cT\sum_{n=-\infty}^{\infty}\int\frac{d{\vec p}}{(2\pi)^3}~\frac{\Sigma_S(\omega_n,\vec p)}{(\Sigma_S^2(\omega_n,\vec p)+Q_n^2)^2}\left(\Sigma_S(\omega_n,\vec p)-\frac{Q_n^2}{2}~\frac{d\Sigma_S(\omega_n,\vec p)}{dQ_n^2}\right),\nonumber
\end{eqnarray}
where $Q_n^2=\omega_n^2+{\vec p}^2$. The behavior of $f_{\pi}(T)$ is shown in Fig.5.

\subsection{Critical Exponents}

Since we have a second order phase transition, we can estimate the critical exponents. To estimate the critical exponents we define the reduced temperature $\tau$ and exponents $\beta$, $\gamma$ and $\beta'$ as follows:
\begin{eqnarray}
\tau&=&\frac{T_c-T}{T_c}~~~(T<T_c),\nonumber\\
|\langle {\bar q}q\rangle_T | &\sim& \tau^{\beta},\nonumber\\
|a_2(T)| &\sim& \tau^{\gamma},\nonumber\\
f_{\pi}(T) &\sim& \tau^{\beta'}.\label{eqn:exp}
\end{eqnarray}

In the spin system the exponent $\gamma$ is defined for the magnetic susceptibility $\chi(T)$. In case of the chiral phase transition, $\chi(T)$ is defined as
\begin{eqnarray}
\chi(T)=\frac{\partial \barq_T}{\partial m_0}\bigg{|}_{m_0=0},\nonumber
\end{eqnarray}
where $m_0$ is the current quark mass. Our exponent $\gamma$ agrees with that of the magnetic susceptibility, if the effect of the finite current mass near $T_c$ is represented as  
\begin{eqnarray}
V(\sigma,m_0,T) \sim a_2(T)\sigma^2+b(T)m_0\sigma,\nonumber
\end{eqnarray}
and, furthermore, if $b(T)$ does not have a singular behavior. 

In Landau theory of second order phase transition, the assumption that $\gamma=1$ leads to $\beta=1/2$. In Ref.\cite{BAR2} they study the critical exponents within the Landau theory. In general $\gamma$ deviates from 1. So we first examine the critical exponent $\gamma$ for three cases of the infrared regularization parameter $t_R=0.05,0.1,0.15$. We determine the exponent $\gamma$ from a slope of the linear log fit
\begin{eqnarray}
\ln |a_2(T)|=\gamma\ln \tau+C,\nonumber
\end{eqnarray}
where $C$ is independent of $\tau$. In order to extract the exponent we use the $\chi^2$ fitting(see Fig.6) and have
\begin{eqnarray}
   \gamma\,=\,\left(
                 \begin{array}{c}
                  1.032 \pm 0.006\\
                  1.008 \pm 0.001\\
                  1.041 \pm 0.008 
                 \end{array}
                 \right) &\;&
   \mbox{for}~t_R\,=\,\left(
               \begin{array}{c}
                0.05\\
                0.10\\
                0.15
               \end{array}
               \right)\nonumber
\end{eqnarray}
These results are consistent with the Landau theory.

In the same way(see Fig.7) we determine the exponents $\beta$ for $t_R=0.1$ and have
\begin{eqnarray}
\beta=0.455 \pm 0.008, \nonumber
\end{eqnarray}
which is appreciably different from the Landau theory.

Next, let us estimate $\beta'$.
We used the Pagels-Stoker formula and Eq.(\ref{eqn:fq}) for calculating $f_{\pi}(T)$, thus we have
\begin{eqnarray}
f_{\pi}(T)=|\barq_T|G(\barq_T,T).\nonumber
\end{eqnarray}
We have checked numerically that $G(\barq_T,T)$ is free from the singular behavior in the limit $\tau \rightarrow 0$. It follows that the behavior of $f_{\pi}(T)$ near $T_c$ is the same as that of the condensate:
\begin{eqnarray}
f_{\pi}(T) \sim \tau^{\beta}.
\end{eqnarray}
Thus we have $\beta'=\beta=0.455$.
The same analysis can be repeated for the exponent of the mass gap $\Sigma_S(n=0,\vec{p}=0)$. 
As far as we use Eq.(\ref{eqn:fq}) as a trial mass function, the exponent of the mass gap agrees with the exponent $\beta$.

Note that we use Eq.(\ref{eqn:exp}) to evaluate the critical exponents.
As a rule, for $D>4$ ($D=$ dimension of space-time) one obtains the universal prediction of the mean field theory. However, for $D<4$ the infrared divergences which appears in the coefficients of the power series expansion invalidate the predictions of the mean field theory\cite{JZJ}. Here we are interested in the case of $D=4$, which is critical. 
Although our results are well fitted by Eq.(\ref{eqn:exp}), there may be logarithmic corrections to the power behavior assumed in Eq.(\ref{eqn:exp}).

\section{SUMMARY AND DISCUSSION}

In summary, we studied the chiral phase transition at finite temperature in the QCD-like gauge theory.
We made use of the CJT effective potential in the improved ladder approximation and took account of the logarithmic behavior in the quark dynamical mass function.
In QCD-like gauge theory it has been pointed out that at zero temperature the physical quantities such as $\barq$, $f_{\pi}$ and $m$ are quite stable under a change of the infrared cutoff. We confirmed that the temperature dependence of these quantities are also stable under a change of $t_R$(see also Ref.\cite{SD}).
Then we put $t_R=0.1$ and investigated the temperature dependence of the effective potential and found the second order phase transition at $T_c=0.185\Lambda_{QCD}$. 
We fixed the value of the pion decay constant $f_\pi$ to 93MeV at zero temperature and had $\Lambda_{QCD}=737$MeV, therefore we obtained $T_c=136$MeV. We also examined the temperature dependence of $\barq_T$ and $f_\pi(T)$.

As we have the second order phase transition, we calculated the critical exponents $\gamma$ and $\beta$ which are the exponents of the quadratic term of the effective potential and the order parameter, respectively.  
We examined $\gamma$ for three cases of the infrared regularization parameter and found $\gamma \sim 1$. This result is consistent with the Landau theory.
However we found that $\beta(=0.455 \pm 0.008)$ appreciably deviates from the value of the Landau theory $1/2$. 
Our approach is somewhat related to the mean field approximation.
However we have introduced the logarithmic behavior in the quark dynamical mass function.
This may be one of the main reasons that $\beta$ deviates from the Landau theory. Our results are different from $O(4)$ Heisenberg model.
We also calculated the exponents of $f_{\pi}(T)$ and $\Sigma_S(n=0,\vec{p}=0)$ and they coincided with $\beta$.

In calculating the temperature dependence of several quantities related to the chiral phase transition and estimating the critical exponents, the effective potential and/or the Schwinger-Dyson equation with a variety of approximation have been used[9,10,11,12]. 
In Ref.\cite{BAR2} they have used a variety of CJT effective potential, which is modified by using Schwinger-Dyson equation in Eq.(\ref{eqn:vg}), and the trial mass function without logarithmic behavior. However they have discussed the critical phenomena within a Landau theory.
In Ref.\cite{ALK} they have used the Coulomb gauge Schwinger-Dyson equation with an instantaneous approximation and have found $\beta=0.49$.
In recent work based on the confining Schwinger-Dyson equation\cite{CDR}, it has been found that $\beta=0.46 \pm 0.04$.
On the other hand, in Ref.\cite{SD} they have used the Schwinger-Dyson equation in the improved ladder approximation and have found that $\beta=0.171$.
Although their model is in the same framework as ours, however, this value largely deviates from the Landau theory and our result.
We calculate the effective potential by using the trial mass function, while the authors of Ref.\cite{SD} solved the Schwinger-Dyson equation.
Furthermore, there are important differences in the handling of the gluon propagator at finite temperature and in the choice of the argument of the running coupling, they used $\bar{g}^2(p,q)=\bar{g}^2(p^2+q^2)$.

In this paper, we have obtained a large value of $\Lambda_{QCD}$. This may be partly due to the fact that we did not consider the effect of the color confinement.
In fact, in Ref.\cite{SASAKI}, they have used an observed value of $\Lambda_{QCD}$ to reproduce the pion decay constant.

As mentioned in Sec.IV, some critical exponents are defined in relation to the current quark mass.
Therefore, it is very interesting to generalize our model to the case of a finite current mass and to study the critical exponents and the scaling relations.

%
%    references
%

\newpage

\begin{center}
{\bf Figure Captions}
\end{center}

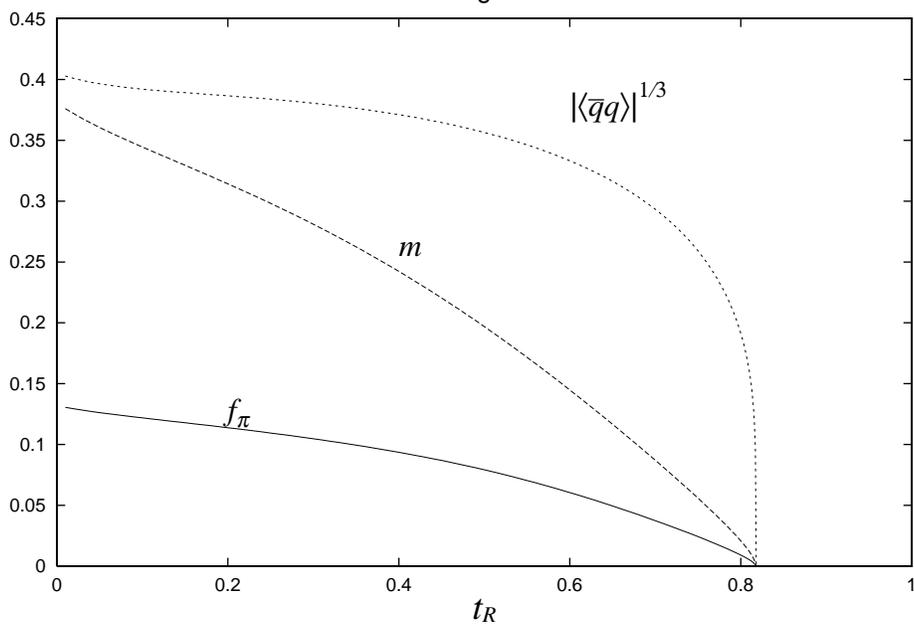
\noindent {\bf Fig.~1:}$t_R$ dependence of $f_{\pi}$, $|\langle\bar{q}q\rangle|^{1/3}, m$ at $T=0$. The mass unit is $\Lambda_{QCD}$.  \\
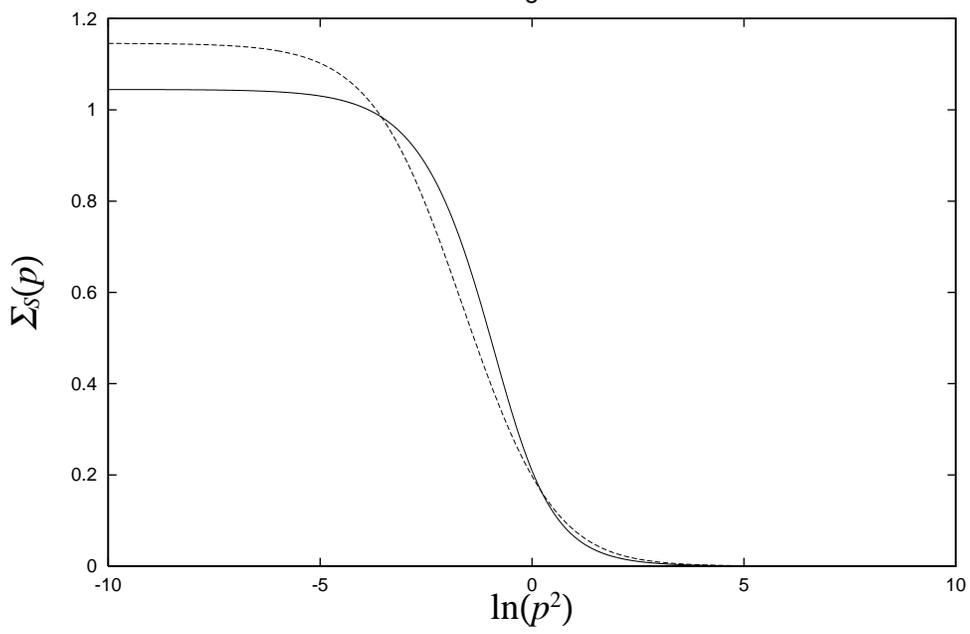
{\bf Fig.~2:}The dynamical mass function of quarks for $t_R=0.1,T=0$. The curves correspond to the solution of Schwinger-Dyson equation (solid line) and the trial mass function Eq.(15) at the minimum of $V$ (dashed line).\\
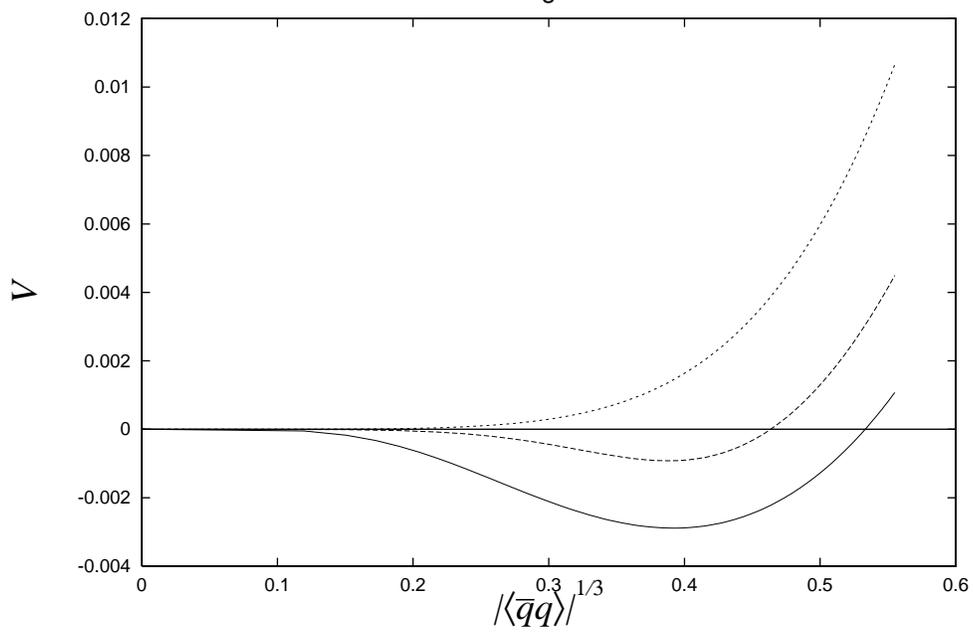
{\bf Fig.~3:}Effective potential at finite temperature as a function of $|\langle\bar{q}q\rangle|^{1/3}$ for case $t_R=0.1$, $N_f=2$ and $N_c=n_f=3$. The curves show the cases $T=0$ (solid line), $T=0.15(\Lambda_{QCD})$ (dashed line), $T=0.25(\Lambda_{QCD})$ (dotted line).\\
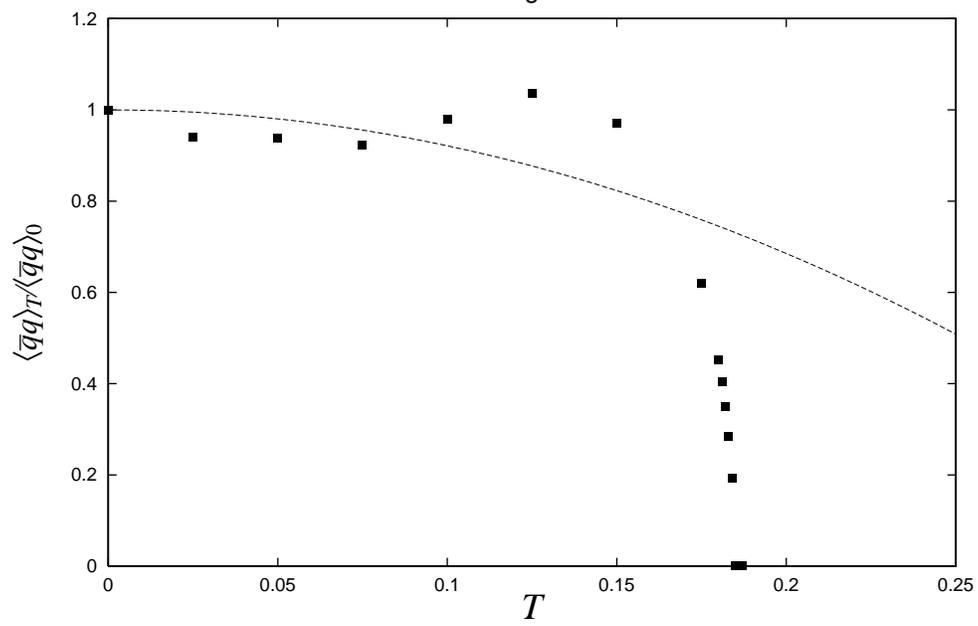
{\bf Fig.~4:}The temperature dependence $|\langle\bar{q}q\rangle_T|$ for $t_R=0.1$. The dashed line corresponds to the free pion gas approximation.\\
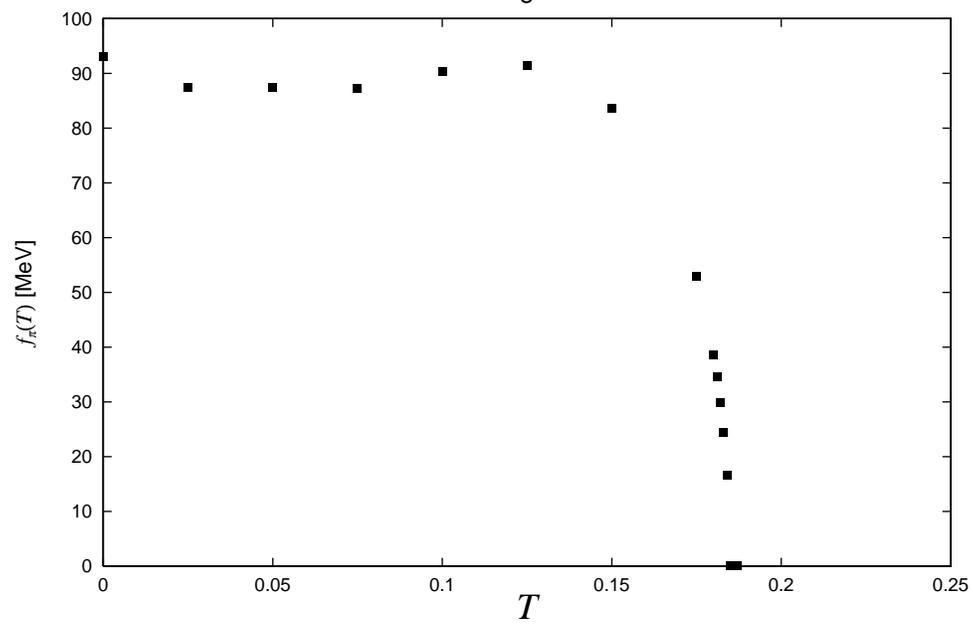
{\bf Fig.~5:}The temperature dependence of $f_{\pi}(T)$ for $t_R=0.1$.\\
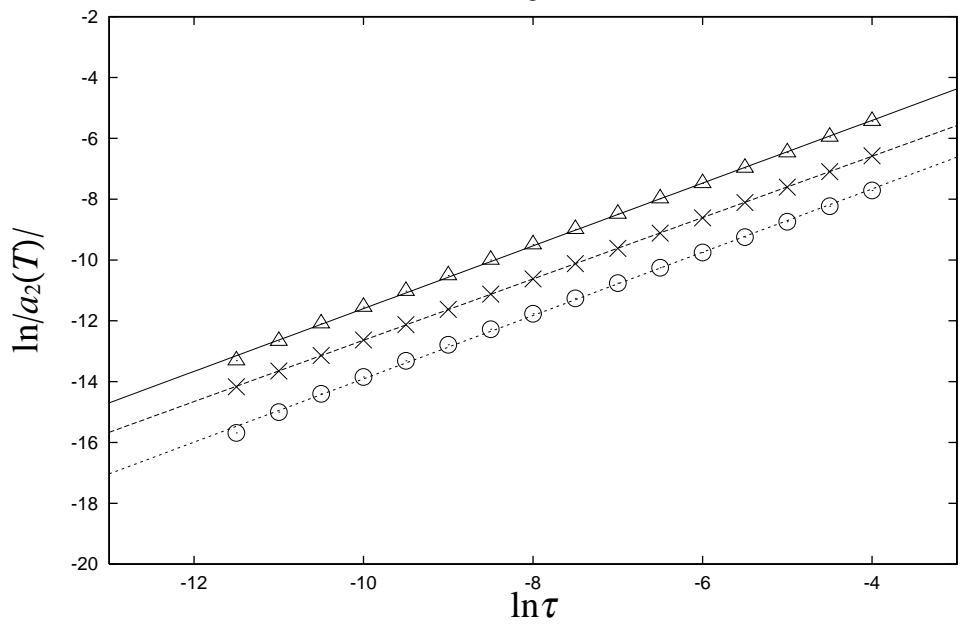
{\bf Fig.~6:}Linear log fit to $\ln |a_2(T)|$ for $t_R=0.05$(triangle), $0.1$(cross), $0.15$(circle) respectively. Here the absolute normalization of $a_2(T)$ is taken arbitrary.\\
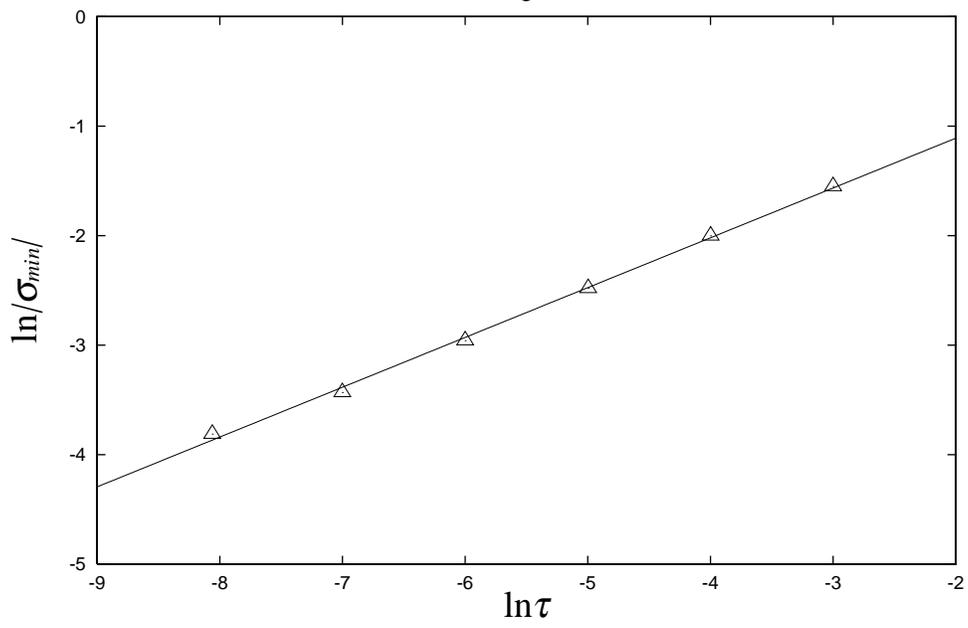
{\bf Fig.~7:}Linear log fit to $\ln\sigma_{min}$ for $t_R=0.1$.

\end{document}